# Magnetostructure of MnAs on GaAs revisited

E. Bauer,[1,2] R. Belkhou,[2,3] S. Cherifi,[4] A. Locatelli,[2] A. Pavlovska,[1,2] and N. Rougemaille[2,4]

[1]Department of Physics, Arizona State University, Tempe, Arizona 85287-1504, USA
[2]Sincrotrone Trieste, S.S.14, Km 163.5, Area Science Park, Basovizza, 34012 Trieste, Italy
[3]Synchrotron SOLEIL, L'Orme des Merisiers, B.P. 48 Saint-Aubin, 91192 Gif-sur-Yvette Cedex, France
[4]Institut Néel, CNRS and Université Joseph Fourier, BP 166, F-38042 Grenoble Cedex 9, France

**Abstract**

The ferromagnetic to nonferromagnetic ($\alpha$-$\beta$) phase transition in epitaxial MnAs layers on GaAs(100) is studied by x-ray magnetic circular dichroism and x-ray magnetic linear dichroism photoemission electron microscopy in order to elucidate the nature of the controversial nonferromagnetic state of $\beta$-MnAs. In the coexistence region of the two phases the $\beta$ phase shows a clear XMLD signal characteristic of antiferromagnetism. The nature and the possible causes of the elusiveness of this magnetic state are discussed.

## I. INTRODUCTION

While the ferromagnetism (FM) of the hexagonal $\alpha$ phase of MnAs [Fig. 1(a)] has been firmly established for nearly 100 years,[1] the magnetic state of the orthorhombic $\beta$ phase [Fig. 1(b)], usually described as paramagnetic (PM), has been the subject of considerable controversy. The controversy started with the suggestion of Guillard[2] that the $\beta$ phase is antiferromagnetic (AFM) because of the increase of the susceptibility with temperature, dramatically violating the Curie-Weiss law characteristic for paramagnetism. Subsequent neutron diffraction studies, however, could not detect AFM long range order,[3] thus apparently eliminating Guillard's proposal. In the search for an explanation of the anomalous magnetic behavior a number of proposals have been made. Bean and Rodbell[4] developed a thermodynamic theory within the molecular field approximation involving a strain dependence of the exchange energy with which they could explain the first-order FM-PM $\alpha$-$\beta$ transition. Goodenough et al.[5] introduced a model in which the crystal changed from a high spin state in the $\alpha$ phase to a low spin state in the $\beta$ phase which allowed them to give a qualitative explanation of the temperature dependence of the susceptibility. Earlier Kittel[6] had already suggested an exchange inversion model in which the abrupt volume change during the phase transition is connected with a change of the sign of the exchange constant via exchange magnetostriction. None of these models could explain all aspects of the two phases and of the transition between them. For example, neutron diffraction studies showed that the $\beta$ phase is not a low spin phase.[7] Other neutron diffraction data indicated that there are spin correlations within distances of less than 20 Å (Ref. 8) but not with AF coupling between nearest neighbors as suggested earlier.[9] Recent polarized neutron scattering experiments[10] found magnetic correlations which indicate ferromagnetic short range order in the $\beta$ phase and also in the high temperature paramagnetic hexagonal $\gamma$ phase. Nevertheless, it was widely accepted that the $\beta$ phase is paramagnetic. All these studies were made on bulk material.

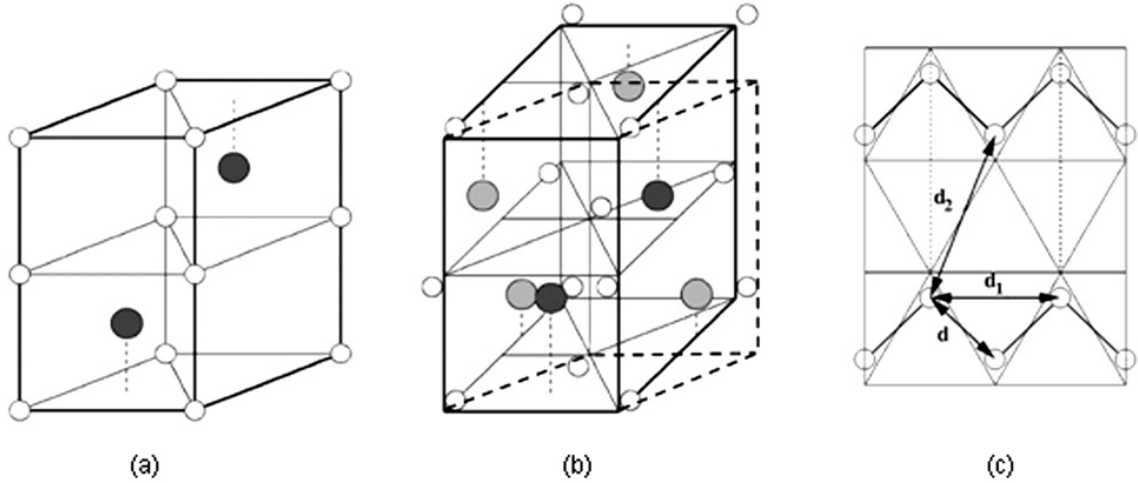

FIG. 1. Structure of MnAs. (a) Hexagonal $\alpha$ phase (NiAs type). (b) Orthorhombic $\beta$ phase (MnP type); the corresponding unit cell dimensions of the $\alpha$ phase are indicated by dashed lines. (c) Mn atomic positions in the $\beta$ phase and interatomic distances in the basal plane. The zigzag chains discussed in the text are indicated by thicker lines. Open circles, Mn; black circles, As in the unit cell; and gray circles, As atoms in neighboring unit cells.

A new phase in the study of MnAs started when Tanaka et al.[11] succeeded to grow epitaxial MnAs layers on GaAs(100). Soon other groups also grew and studied these layers and one of them[12] made an important observation: when the films were cooled in a magnetic field ("field cooled") from high temperatures below the $\alpha$-$\beta$ transition temperature their hysteresis curve was asymmetric, i.e., it showed an exchange bias typical for FM/AFM interfaces. This was attributed to strain-stabilized AFM $\beta$-phase inclusions in the $\alpha$ phase. The exchange bias field $H_E$ which is the measure of the asymmetry of the hysteresis curve approached a constant value at low temperature but went rapidly to zero in the $\alpha$-$\beta$ transition region. The exchange bias in field-cooled layers was recently confirmed again at room temperature.[13] Most of the work on the epitaxial MnAs film, however, was concentrated on the ferromagnetic $\alpha$ phase, which is now well understood, mainly due to the many multimethod studies of the Däweritz/Ploog group as documented in an extensive recent review by Däweritz.[14] In nearly all of this work including a very recent study[15] the $\beta$ phase was considered to be paramagnetic. Only recently, stimulated by the theoretical work to be discussed next, experiments have been interpreted as supporting an AFM $\beta$ phase.[16,17]

Although there have been a number of band structure calculations addressing the problem of the magnetic state of the β phase, only the most recent and relevant ones will be considered here. They use different methods and models and come to different conclusions, thus keeping the discussion of the β phase alive. The first one[18] compares three AFM phases with different spin distributions with the FM and PM phases and comes to the conclusion that a random arrangement of AFM planes has the lowest energy for the lattice parameters of the β phase. The second[19] one considers the same AFM phases and the FM and PM phases and finds that the PM phase has the lowest energy. The PM phase is also favored by the third one,[20] which uses a quite different approach. Thus, at present it appears that from theory the β phase is most likely paramagnetic, at least in the bulk. This does not exclude an AFM state in epitaxial layers that are highly strained so that exchange magnetostriction could favor the AFM coupling, as suggested by Kittel[6] long ago.

## II. EXPERIMENT

The most direct way to search for AFM is x-ray magnetic linear dichroism (XMLD) spectroscopy. In the case where FM and AFM regions coexist a microscopic technique is useful in order to separate the signals from the two regions. This is achieved by XMLD photoelectron emission microscopy (XMLDPEEM). Linear dichroism (LD) is proportional to the quadrupole moment of the charge distribution which in AFM materials depends on the direction of the magnetic axis via the spin-orbit coupling. In this case the magnetic contribution to the LD signal is proportional to $3(\bar{A} \cdot P)-1$, where $\bar{A}$ and $P$ are unit vectors in the directions of the magnetic axis and of the electric field, respectively. In XMLDPEEM $2p$ core electrons are excited by the incident linear polarized light from a synchrotron radiation source into empty $d$ states. The secondary electrons produced in the filling process of the core hole are used then for imaging. By rotating $P$ the direction of $\bar{A}$ can be determined. The $2p$ ($L_3, L_2$) absorption edges have a structure due to final state multiplet coupling between the $2p$ core hole and the $3d$ valence band hole, as illustrated in Fig. 2 for the $L_3$ edge of Mn in MnAs. The XMLDPEEM images shown below (Figs. 5 and 6) are difference images between images taken from the two energies indicated in Fig. 2, normalized by the sum of these images.

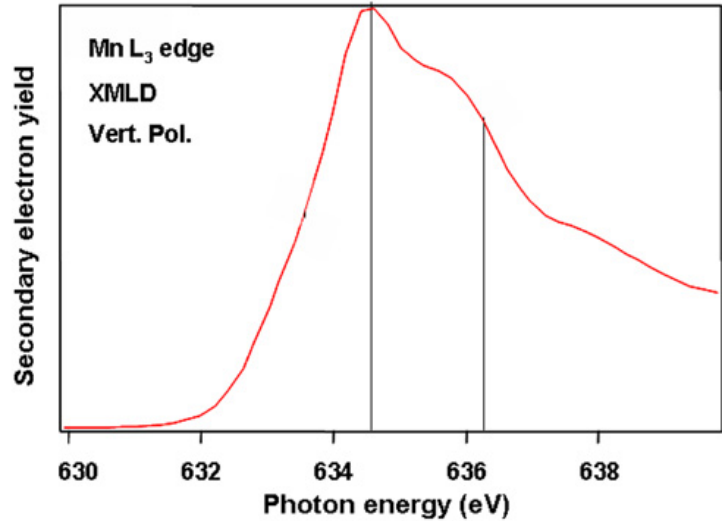

FIG. 2. Mn $L_3$ ($2p_{3/2}$) absorption spectrum. The secondary electron yield is plotted as a function of the exciting photon energy. The lines indicate the photon energy used for the two images whose difference is displayed in the XMLDPEEM images (Figs. 5 and 6).

The ferromagnetic regions are imaged by x-ray magnetic circular dichroism photoelectron emission electron microscopy (XMCDPEEM). Here also the secondary electrons produced during the filling of the $2p$ core hole are used for imaging but the contrast mechanism is quite different. It is based on the difference in the density of unfilled $d$ states available for the photoexcited $2p$ core electron. Depending on the helicity (right or left circular polarization) of the incident light, transitions into the minority or majority states are dominant, resulting in a corresponding difference in the secondary electron emission.

In the present study both magnetic imaging methods were combined with low energy electron microscopy (LEEM) and low energy electron diffraction for structural characterization. More about these imaging techniques can be found in the corresponding chapters in Ref. 21. The experiments were performed in the two spectroscopic photoemission LEEM (SPELEEM) systems at the nanospectroscopy beamline of the synchrotron radiation facility ELETTRA in Trieste, Italy. One of the instruments, which has a 60° beam separator, is described in Ref. 22; the other one has a 90° beam separator but is otherwise similar. The properties of the light source can be found in Ref. 23.

The samples were prepared in the laboratory of Däweritz by solid-source molecular beam epitaxy at 200-250 °C in the so-called A orientation in which the MnAs (1−100) plane is parallel to the substrate surface and the MnAs [11–20] direction parallel to the GaAs [110] direction. The films were annealed at 400 °C and capped with As before removal from the vacuum system for transfer to the SPELEEM in which they were decapped. Numerous films with different thicknesses and orientations have been studied with XMCDPEEM after the first experiments.[24] Some of the results, which have contributed considerably to the understanding of the ferromagnetic $\alpha$ phase, are described in Ref. 14 and references therein. The XMCDPEEM studies reported here were done only at two thicknesses, 300 and 120 nm in the 90° beam separator SPELEEM instrument.

## III. RESULTS

We begin with a brief overview of some of the results from our previous XMCDPEEM studies of the $\alpha$ phase that are relevant for the present study. In the coexistence range of the two phases the strain relaxation causes the formation of stripes whose periodicity depends linearly on thickness in the thickness range studied (20-500 nm) and whose relative width depends on temperature. Figure 3 shows an example from the intermediate thickness range. The black and white regions are FM with opposite magnetization perpendicular to the stripe direction; the gray regions give no FM signal and are supposed to be "PM." In the thinnest films, which have a simple FM domain structure, the $\beta$ fraction changes approximately linearly with temperature as, shown in Fig. 4 for a 40 nm thick film. In thicker films, in which the domain structure is more complicated, the change is also more complicated, as shown elsewhere.[24]

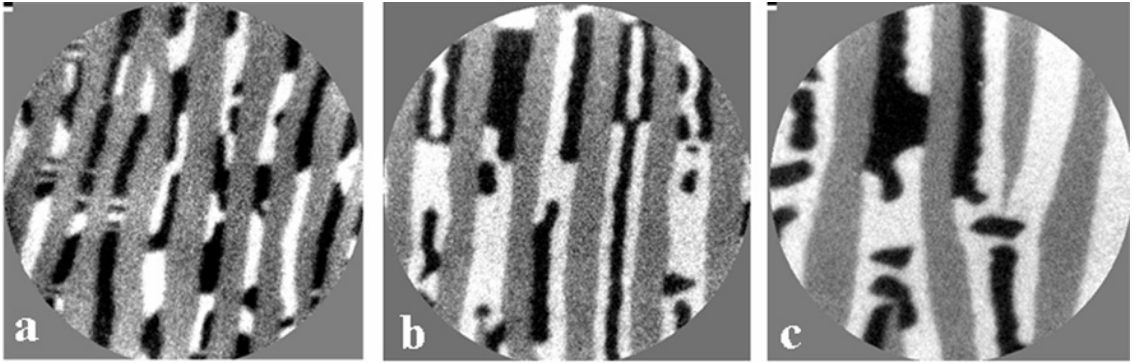

FIG. 3. XMCDPEEM images from MnAs on GaAs(100) illustrating the dependence of the stripe period and of the magnetic domain structure at room temperature upon film thickness; (a) 120 nm, (b) 180 nm, and (c) 300 nm. Field of view is 5 μm.

In the present study the samples could not be cooled. Thus, only experiments from room temperature upwards could be made. Figure 5 compares the XMCDPEEM and XMLDPEEM images of 300 and 120 nm thick films. The comparison of the right and left images clearly shows the 1:1 correlation between the bright AFM stripes (right) and the zero intensity stripes between the FM stripes (left). The XMLDPEEM images were taken with vertical polarization, i.e., with the **E** vector of the light, which has a glancing angle of incidence on the sample, nearly perpendicular to the sample surface. The signal seen in the bright "PM" stripes, therefore, is a measure for the out-of-plane component of the AFM magnetic axis $\bar{\mathbf{A}}$. A similar but weaker signal is obtained for horizontal polarization, which is a measure for the in-plane component of $\bar{\mathbf{A}}$. This means that $\bar{\mathbf{A}}$ is neither in-plane nor perpendicular to the surface but tilted and closer to the film normal.

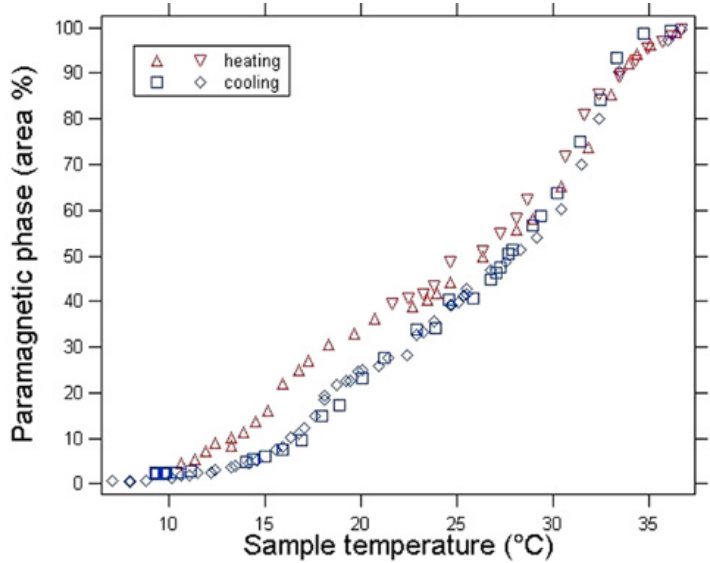

FIG. 4. Dependence of the "paramagnetic" fraction of a 40 nm thick MnAs layer on GaAs upon temperature. The data were taken from LEEM images in which the stripe structure produces a phase contrast due to the height differences between ferromagnetic and paramagnetic stripes.

In the images shown in Fig. 5 the plane of incidence of the photon beam was normal to the stripe direction, which gives only partial information on the direction of $\bar{\mathbf{A}}$. Its full characterization can be obtained by rotating the sample 90°. This was done with the 120 nm thick film. Analysis of all four component images gave tilt angles of $\bar{\mathbf{A}}$ of about 40° to the film normal and to the magnetic easy axis of the FM stripes ([11–20]). The small XMLD signal and signal-to-noise level in the difference images of the 120 nm thick film cause a large uncertainty in these values, so that they should be considered as preliminary.

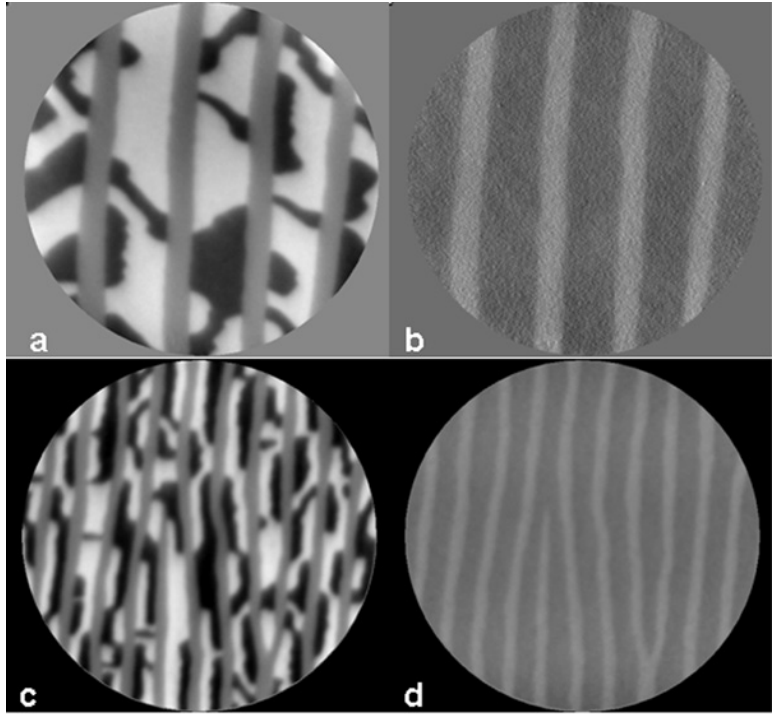

FIG. 5. XMCDPEEM (left) and XMLDPEEM (right) images of 300 nm (top) and 120 nm (bottom) thick MnAs layers on GaAs(100) at room temperature. Field of view is 5 μm.

The temperature dependence of the AFM signal was studied in both films and is illustrated in the XMLDPEEM images of Fig. 6 for the 300 nm thick film, together with the corresponding XMCDPEEM images. Again the correlation between the images on top and on the bottom is clearly seen. In the XMLDPEEM images the intensity in the AFM stripes decreases with temperature simultaneously with the decrease of the width and contrast of the FM stripes, i.e., with the magnetic field acting on them. Again, the low signal-to-noise ratio causes large error bars, but to a first approximation the AFM signal decreases linearly and disappears together with the FM regions at the $\alpha$-$\beta$ transition temperature. The temperatures shown in Fig. 6 have a small systematic error and should be considered only as relative values. It is interesting to note that some very weak stripe contrast can be seen at much higher temperatures (about 100 °C) which is not understood at present and needs further study. It should be noted that the films were not exposed to an external magnetic field during cooling so that the "PM" stripes were only in the field of the FM stripes which is predominantly in-plane perpendicular to the stripes.

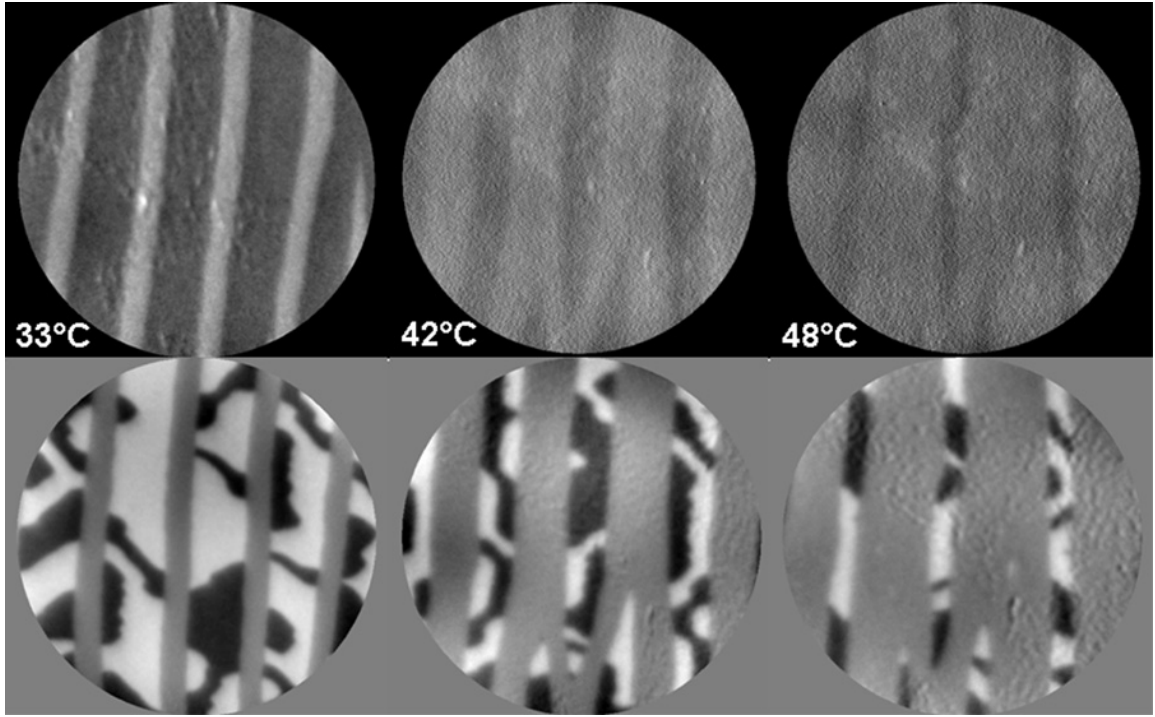

FIG. 6. Temperature dependence of the AFM (top) and of the FM (bottom) contrasts and fractions in a 300 nm thick film. The temperatures are from left to right 33, 42, and 48 °C. Field of view is 5 μm.

## IV. DISCUSSION

The XMCDPEEM results reported in the previous section clearly show that the "PM" stripes in the coexistence region of the $\beta$ and $\alpha$ phases are AFM with a tilted magnetic axis, even in the absence of an applied magnetic field during cooling. However, the observation that a clear AFM signal is seen only in the presence of the $\alpha$ phase suggests that the PM phase becomes AFM only in the presence of a magnetic field. The magnetization studies in which an exchange bias was observed only upon field cooling but not in the absence of an applied field[12,13] point in the same direction. The apparent discrepancy between the two types of studies can be explained as follows. Without an applied field during cooling the two sides of the FM stripes have with equal probability the magnetization pointing to the right and to the left, as illustrated in the XMCDPEEM images shown in Fig. 7. Accordingly the alignment of the AFM regions at the FM/AFM boundary induced by the magnetic field of the FM stripes will also be equally probable in both directions. As a result the asymmetry of the interface pinning, which is necessary for exchange bias to occur, will be lost. During cooling in an applied field all FM stripes are magnetized in the same direction and as a consequence also all AFM regions at the FM border, producing the necessary asymmetry. In XMLDPEEM no interface pinning is needed, only a preferred direction of the AFM domains.

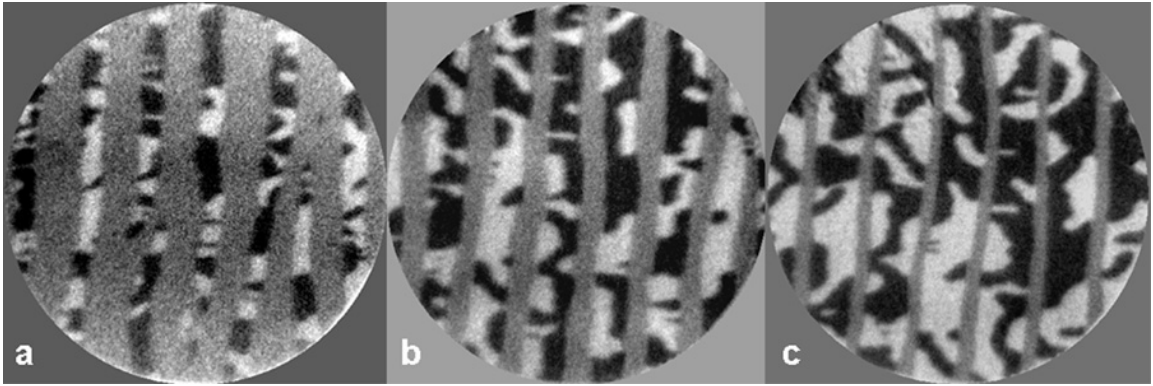

FIG. 7. Temperature dependence of the domain structure of a 180 nm thick MnAs film on GaAs(311) during cooling from left to right. The same irregular domain pattern is observed on GaAs(100). Field of view is 5 μm.

The fact that these domains must be small is evident from the absence of long range order as determined by neutron diffraction. The work in Ref. 9 suggests domain sizes of the order of 10 Å. The spin structure within these regions and their boundaries can be postulated at present only on the basis of the recent density functional calculations.[18,19] Both calculations consider the same three spin configurations for the AFM phase. Counting the two Mn atoms in the basal plane of the β-phase unit cell as 1,2 and the corresponding two Mn atoms in its midplane as 3,4 (see Fig. 1), the spin configurations are (A) ↑↑↓↓, (B) ↑↓↑↓, and (C) ↑↓↓↑. All three AFM configurations have a somewhat higher energy $E$ (lower binding energy) than the FM configuration (↑↑↑↑) in their respective optimized structure. Both calculations clearly favor in-plane AFM coupling over in-plane FM coupling and differ only in the energies of the two in-plane coupling configurations: $E_C > E_B > E_{FM} + 45$ meV in Ref. 18 while $E_C > 2E_B = E_{FM} + 35$ meV in Ref. 19, slightly favoring configuration B with AFM coupling within the basal plane and FM coupling perpendicular to it.

However, the thermal energy at 350 K is 30 meV, so that configuration C with AFM coupling in both directions has also a finite probability.

From these data the following picture emerges of the AFM structure if it exists: it consists of domains with spin configuration B separated from each other by domain walls with spin configuration C which may be considered as stacking faults of configuration B. This is similar to the disorder proposed in Ref. 18. As far as the "if" is concerned it must be pointed out that Ref. 18 concludes that $\beta$-MnAs is AFM, based on the considerably higher energy of the PM phase, while Ref. 19 concludes that $\beta$-MnAs is PM, based on the result that at the experimental lattice dimensions $E_{AFM}$ is considerably larger than $E_{FM}$ and comparable to $E_{PM}$ which has a minimum at these dimensions. The AFM configuration is favored only at lattice distortions that are considerably larger than the experimentally observed one. The discrepancy between the two conclusions has been attributed to the assumption of zero magnetic moment of the Mn atoms in the PM state in Ref. 18.[25] Thus from the point of the theory, $\beta$-MnAs must be considered to be PM at present, at least in the bulk. It could be argued that the large strain in epitaxial layers on GaAs could be the cause of AFM. However, precise synchrotron x-ray diffraction experiments[26] have shown that the misfit strain is relaxed by misfit dislocations and that it decays within a very thin interface layer. It certainly cannot cause AFM in a 300 nm thin film. Misfit strain as the cause of AFM is also excluded by the measurements of the exchange bias[12] which was found to be negligible in the thinnest films.

One possible explanation that could reconcile present theory and experiment is that the positions of the Mn atoms in the basal plane in the epitaxial films are different from those in the bulk. The calculations of Ref. 19 show that the main driving force towards AFM is the exchange coupling $J$ between inequivalent Mn atoms within the Mn zigzag chains with distance $d$ [see Fig. 1(c)]. While the exchange coupling between neighboring basal plane layers and that between equivalent in-plane Mn atoms $J_1$ changes little with distortion and remains FM with increasing distortion (because the atomic distances change only little), $J$ becomes AFM already at 75% distortion, where 100% distortion corresponds to equilibrium distortion of the $\beta$ phase. If the distortion in epitaxial layers should be larger than in the bulk, AFM may be stabilized in them even if the bulk is PM. A larger distortion could possibly be induced by thermal strain. Film and substrate have different expansion coefficients, in particular, along the **b** axis of $\beta$-MnAs. Upon cooling from the growth temperature (250 °C) to room temperature thermal stress-induced shifts of the Mn positions along the **b** axis [Fig. 1(c)] could occur. However, they would have to be rather large in order to produce the lattice distortion needed for AFM to be favored according to Ref. 19. In thick films the thermal stress is partially relaxed by cracks perpendicular to the **b** axis.

Another possibility for reconciliation of theory and experiment which does not involve different Mn basal plane positions in the bulk and in epitaxial layers is the fact that theory does not take into account lattice vibrations. Their importance is illustrated by the observation that anomalous changes in the lattice dynamics, which show strongly anisotropic vibration amplitudes with maximum amplitude approximately perpendicular to the zigzag chains, are more important than the static changes considered in the calculations.[27] Because the coupling constant $J$ between the nearest neighbors in the chains is important for AFM coupling, these vibrations should have a significant influence on the coupling. Kittel[6] has already pointed out that the indirect spin-spin coupling via phonons plays an important role when the magnetocrystalline anisotropy is high, which is the case in MnAs.

The disappearance of AFM upon the transition to the high temperature $\gamma$ phase may be caused by the breakup of the magnetic coupling by thermal fluctuations because the thermal energy at the transition temperature (110 meV) is significantly larger than the coupling constants. This may be aided by the proliferation of magnetic domain walls, causing the disintegration of the domains. The breakup of the coupling is accompanied by the transition from anisotropic to isotropic vibrations,[27] which is an indication of the intimate relationship between spin-spin and spin-phonon couplings.

## V. SUMMARY

We have shown unambiguously with XMLDPEEM that the $\beta$ phase of epitaxial MnAs layers on GaAs(100) is antiferromagnetic at least in the $\alpha$-$\beta$ coexistence region, even in the absence of an external applied field during cooling. A comparison with the literature suggests that an applied field during cooling leads to a uniaxial alignment of nanoscopic antiferromagnetic domains in which the Mn atoms are coupled antiferromagnetically within the basal plane and ferromagnetically perpendicular to it. Antiferromagnetism is present only on the nanoscale so that the $\beta$ phase may also be considered to be superparamagnetic similar to the state of nanoscale ferromagnets. There are still many open questions, in particular, concerning the magnetic state between the α-β and the β-γ transitions in the absence of long range ferromagnetic order, concerning the size of the magnetically ordered regions and concerning the details of the influence of an applied field during cooling. Future x-ray magnetic dichroism and polarized x-ray scattering studies without and with an applied field and an improved polarized neutron scattering work over a wider temperature range will hopefully answer these questions.

ACKNOWLEDGMENTS

One of the authors (E.B.) wants to thank L. Däweritz for providing the samples and S. Sanvito for clarifying correspondence. He also acknowledges support by Office of Naval Research under Grant No. N-000140210922 and by NATO.

**References**

[1] S. Hilpert and T. Diekmann, Ber. Deutsche Chem. Gesellsch. A **44**, 2378 (1911).

[2] C. Guillard, J. Phys. Radium **12**, 223 (1951).

[3] G. E. Bacon and R. Street, Nature (London) **175**, 518 (1955).

[4] C. P. Bean and D. S. Rodbell, Phys. Rev. **126**, 104 (1962).

[5] J. B. Goodenough, D. H. Ridgley, and W. A. Newman, Proceedings of the International Conference of Magnetism, Nottingham, 1964, p. 542.

[6] C. Kittel, Phys. Rev. **120**, 335 (1960).

[7] L. H. Schwartz, E. L. Hall, and G. P. Felcher, J. Appl. Phys. **42**, 1621 (1971).

[8] N. N. Sirota and G. A. Govor, Phys. Status Solidi B **43**, K165 (1971).

[9] N. P. Grazhdankina, E. A. Zavadskii, and I. G. Zakidov, Sov. Phys. Solid State **11**, 1879 (1969).

[10] K.-U. Neumann, K. R. A. Ziebeck, F. Jewiss, L. Däweritz, K. H. Ploog, and A. Murani, Physica B **335**, 34 (2003).

[11] M. Tanaka, J. P. Harbison, T. Sands, T. L. Cheeks, V. G. Keramides, and G. M. Rothberg, J. Vac. Sci. Technol. B **12**, 1091 (1994).

[12]S. H. Chun, S. J. Potashnik, K. C. Ku, J. J. Berry, P. Schiffer, and N. Samarth, Appl. Phys. Lett. **78**, 2530 (2001).

[13]K.-S. Ryu, D.-H. Kim, S.-C. Shin, and H. Akinaga, Phys. Rev. B **71**, 155308 (2005).

[14]L. Däweritz, Rep. Prog. Phys. **69**, 2581 (2006).

[15]L. B. Steren, J. Milano, V. Garcia, M. Marangolo, M. Eddrief, and V. H. Etgens, Phys. Rev. B **74**, 144402 (2006).

[16]H. Yamaguchi, A. K. Das, A. Ney, T. Hesjedal, C. Pampuch, D. M. Schaadt, and R. Koch, Europhys. Lett. **72**, 479 (2005).

[17]A. Ney, T. Hesjedal, and K. H. Ploog, Phys. Rev. B **72**, 212412 (2005).

[18]M. K. Niranjan, B. R. Sahu, and L. Kleinman, Phys. Rev. B **70**, 180406 (2004).

[19]I. Runger and S. Sanvito, Phys. Rev. B **74**, 024429 (2006).

[20]L. M. Sandratskii and E. Şaşıoğlu, Phys. Rev. B **74**, 214422 (2006).

[21]*Science of Microscopy*, edited by P. W. Hawkes and J. C. H. Spence. (Springer, New York, 2007), pp. 605 and 657.

[22]*High-Resolution Imaging and Spectroscopy of Materials*, edited by F. Ernst and M. Rühle (Springer, Berlin, 2003), pp. 363.

[23]A. Locatelli, D. Cocco, S. Cherifi, S. Heun, M. Marsi, M. Pasqualetto, and E. Bauer, J. Phys. IV **104**, 99 (2003).

[24]E. Bauer, S. Cherifi, L. Däweritz, M. Kestner, S. Heun, and A. Locatelli, J. Vac. Sci. Technol. B **20**, 2539 (2002).

[25]S. Sanvito (private communication).

[26]D. K. Satapathy, V. M. Kaganer, B. Jennichen, W. Braun, L. Däweritz, and K. H. Ploog, Phys. Rev. B **72**, 155303 (2005).

[27]G. A. Govor, Sov. Phys. Solid State **23**, 841 (1981).